\documentclass[aps,twocolumn,showpacs,superscriptaddress,nofootinbib]{revtex4}

\usepackage{graphicx}
\usepackage{latexsym}

\usepackage{color}

\begin{document}

\title{Process Time Distribution of Driven Polymer Transport}

\author{Takuya Saito}
\email[Electric mail:]{saito@stat.phys.kyushu-u.ac.jp}
\affiliation{Department of Physics, Kyushu University 33, Fukuoka 812-8581, Japan}

\author{Takahiro Sakaue}
\email[Electric mail:]{sakaue@phys.kyushu-u.ac.jp}
\affiliation{Department of Physics, Kyushu University 33, Fukuoka 812-8581, Japan}
\affiliation{PREST, JST, 4-1-8 Honcho Kawaguchi, Saitama 332-0012, Japan}

\def\Vec#1{\mbox{\boldmath $#1$}}
\def\degC{\kern-.2em\r{}\kern-.3em C}

\def\SimIneA{\hspace{0.3em}\raisebox{0.4ex}{$<$}\hspace{-0.75em}\raisebox{-.7ex}{$\sim$}\hspace{0.3em}} 

\def\SimIneB{\hspace{0.3em}\raisebox{0.4ex}{$>$}\hspace{-0.75em}\raisebox{-.7ex}{$\sim$}\hspace{0.3em}}

\date{\today}

\begin{abstract}
We discuss the temporal distribution of dynamic processes in driven polymer transport inherent to flexible chains due to stochastic tension propagation.
The stochasticity originates from the disordered initial configuration of an equilibrium polymer coil, which results in random paths for tension propagation.
We consider the process time for when translocation occurs across a fixed pore and when stretching occurs by pulling the chain end.
A scaling argument for the mean and standard deviation of the process time is provided using the two-phase picture for stochastic propagation. 
The two cases are found to differ remarkably.
The process time distribution of the translocation exhibits substantial spreading even in the long-chain limit, unlike that found for the dynamics of polymer stretching.
In addition, the process time distribution in the driven translocation is shown to have a characteristic asymmetric shape.
\end{abstract}


\def\degC{\kern-.2em\r{}\kern-.3em C}

\newcommand{\gsim}{\hspace{0.3em}\raisebox{0.5ex}{$>$}\hspace{-0.75em}\raisebox{-.7ex}{$\sim$}\hspace{0.3em}} 
\newcommand{\lsim}{\hspace{0.3em}\raisebox{0.5ex}{$<$}\hspace{-0.75em}\raisebox{-.7ex}{$\sim$}\hspace{0.3em}} 

\maketitle

\section{Introduction}
Polymer transport is a ubiquitous and fundamental process in biological science and technology.
A good example is polymer translocation in which a polymer (e.g., DNA or RNA) is driven across a small pore by a chemical or electric potential difference.
Individual translocation events can be detected by applying an external force~\cite{PNAS_Kasianowicz_1996,PRL_Henrickson_2000,PRL_Meller_2001,NanoLett_Storm_2005,PRE_Storm_2005,BJ_Wanunu_2008,NanoLett_Fologea_2005}, allowing the fundamental properties of polymer transport to be investigated.
In addition, many theoretical investigations have been conducted~\cite{PRL_Sung_1996,PRE_Kantor_2004,MacromolSymp_Marzio_2005,PRE_Sakaue_2007,PRE_Sakaue_2010,EPL_Dubbeldam_2007,JPhys_Vocks_Panja_2008,EPJE_Saito_Sakaue_2011,JPCB_Rowghanian_Grosberg_2010,Arxiv_Dubbeldam_2011,arXiv_Ikonen_Sung_2011}.
Statistical results obtained in experiments exhibit a wide distribution in translocation times (i.e., the time taken for a polymer to pass through a pore)~\cite{PRE_Storm_2005,BJ_Wanunu_2008}.
In addition, numerical simulations have generated considerable insight~\cite{JPhys_Milchev_2011,PRE_Luo_2008,EPL_Luo_2009,JCP_Luo_2006,EPL_Lehtola_2009,EPL_Dubbeldam_2007,JPhys_Vocks_Panja_2008,PRE_Bhattacharya_Binder_2010,EPJE_Bhattacharya_Luo_2009,PRE_Fyta_2008,arXiv_Ikonen_Sung_2011},
which also indicated a broad distribution of the translocation time~\cite{EPJE_Bhattacharya_Luo_2009,JCP_Luo_2006}.
Currently, the mean translocation time is attracting a lot of interest, particularly its scaling exponents with respect to chain length $N_0 a$ ($N_0$ is the segment number with size $a$) and driving force $f_{}$. However, fluctuations in the translocation time need to be clarified to fully understand the translocation process, but they have received less attention to date.

As is well known, a polymeric chain in equilibrium is a fluctuating coil whose configuration is mathematically described by the trajectory of a random (or self-avoiding) walk.
It seems reasonable that when a force is suddenly applied to a portion of the chain, the entire chain will not immediately respond to the stimulus, but only the portion of the chain that is initially set in motion.
The responding domain gradually evolves over time with the propagation of tension along the chain backbone.
In domain growth, the propagation front of the tension follows the disordered trajectory characteristics of a random coil configuration.
The present study investigates the stochastic dynamics of tension propagation, which affects the process time distribution of driven polymer transport.

\section{Origin of stochasticity}
There are generally two sources of the stochasticity in stochastic processes: noise (random forces) and uncertainty in the initial distribution. In the case of polymer transport, the former source is the Brownian force, which is thermal in origin, while the latter source is reflected in the equilibrium configuration when one end of the chain arrives at the pore at time $t=0$.
To quantify the relative importance of these two sources, we consider the time scale involved in polymer transport over a distance $X$ by an external force.
Two time scales can be identified: the diffusion time $\tau_{D} (X) \simeq X^2/D$, where $D$ is the diffusion coefficient, which depends on the chain conformation, and the convection time $\tau_{f_{}}(X) \simeq X/V$, where $V = f_{}/\Gamma$ is the mean biased velocity and $\Gamma$ is the frictional coefficient, which is related to the diffusion coefficient through the Einstein relation $D = k_{\rm B}T/\Gamma$. 
The ratio $Pe (X) \equiv \tau_D(X)/\tau_{f_{}}(X) \simeq f_{} X/(k_{\rm B}T)$ defines a Peclet number. From the condition 
$Pe (R_{\rm eq})>1$, where $R_{\rm eq}$ is the equilibrium coil size, the condition for the driven transport regime can be derived as $f_{} R_{\rm eq}/(k_{\rm B}T) \gsim 1$,
where the effects of thermal fluctuations become relatively unimportant. For stronger forces $Pe (a)>1 \Leftrightarrow f_{} a/(k_{\rm B}T) \gsim 1$, the Brownian force can be neglected even when considering the transport of a single segment.
On the other hand, there is inevitably an uncertainty in the initial distribution, irrespective of the driving force. This uncertainty is expected to be a major cause of the wide translocation time distributions observed in experiments.

Below, we neglect minor effects associated with the stochasticity due to the Brownian force.
We consider the transport time distribution due to stochastic tension propagation along a random backbone trajectory for two cases: driven translocation across a pore
for which the external force is spatially fixed and stretching by suddenly pulling one chain end, for which the forcing point is fixed.
The stochastic scenario allows higher cumulants (e.g., the standard deviation and skewness) to be discussed in addition to the average of the process time distribution.

\begin{figure}[t]
\begin{center}
\includegraphics[scale=0.50]{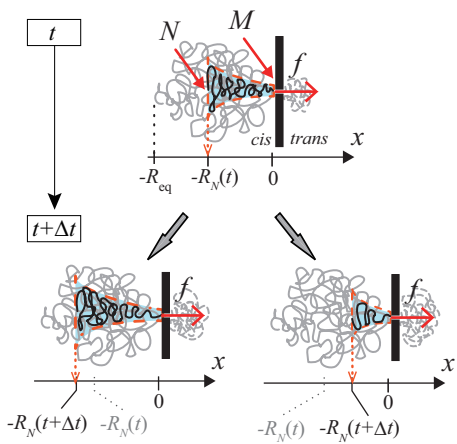}
      \caption{(Color online) Schematic representation of driven polymer translocation (in the trumpet [TP] regime). The tensed moving domain is shaded. The propagation front of the tension is located at $x=-R_N$. The top and bottom snapshots depict stochastic evolution along the initial disordered configuration during time interval $\Delta t$.}
\label{fig1}
\end{center}
\end{figure}

\section{Driven Translocation}
We first consider driven translocation.
To analyze the nonequilibrium response, we adopt the two-phase picture~\cite{PRE_Sakaue_2007,PRE_Sakaue_2010,JPCB_Rowghanian_Grosberg_2010,EPJE_Saito_Sakaue_2011} in which the chain separates on the {\it cis} side into quiescent and moving domains.
As illustrated in Fig.~1, the coordinate $x$ is taken to be perpendicular to the wall with the pore.
The driving force acts only at the pore site ($x=0$) with a constant force magnitude $f_{}$ in the $x$-direction from the {\it cis} to the {\it trans} side.
The linear polymer consisting of $N_0$ segments is initially in $x \leq 0$.
The polymer segments are numbered from one end of the chain to the other end ($N_0$th segment); the first segment arrives at the pore at time $t=0$.
As time evolves, the rear segments are gradually sucked and the moving domain on the {\it cis} side grows in the negative $x$-region.
At time $t$, $M(t)$ is the number of the segment at the pore ($x=0$) and $N(t)$ is the segment number at the tension-propagation front ($x=-R_N$).
The propagation front follows a random path sampled from equilibrium coil configurations.
Consequently, the dynamics of a particular sample strongly depends on how the polymer is brought to the pore.
Below, the length, force, and time are made dimensionless by employing units of the segment length $a$, force $k_{\rm B}T/a$, and time $\eta a^3/(k_{\rm B}T)$, respectively (where $k_{\rm B}T$ is the thermal energy and $\eta$ is the solvent viscosity).

Here, we introduce the relevant exponents. The Flory exponent $\nu$ and the dynamical exponent $z$ are associated with the static and dynamic properties, respectively; the equilibrium coil size is described by $R_{\rm eq} \simeq N_0^{\nu}$ and the hydrodynamic drag force acting on the coil with moving velocity $V$ is described by $\sim R_{\rm eq}^{z-2}V$~\cite{deGennesBook}. 
The exponents have the following values in practical important cases: $\nu=1/2$ (for an ideal chain), $\nu=\nu_3 \simeq 0.5876...$ (for a three-dimensional self-avoiding chain),
$z=3$ (for non-draining case), and $z=(1+2\nu)/\nu$ (for free-draining case).

The moving domain is characterized by the dynamical equations of state, \cite{EPL_Brochard_1993,EPL_Brochard_1995,PRE_Sakaue_2007,PRE_Sakaue_2010,EPJE_Saito_Sakaue_2011}, which describe the global polymer conformation in terms of the driving force $f$, the extension $R_N$ along the $x$-axis, and the representative velocity $V$:

\begin{equation}
N(t)-M(t) = \sigma_F (f) R_N,
\label{N_M_f_R}
\end{equation}
\begin{eqnarray}
R_N V = {\cal Z} (f_{}) 
\label{R_V_general} 
\end{eqnarray}
with
\begin{eqnarray}
\sigma_F(f)
&\simeq& \left\{ \begin{array}{ll}
f_{}^{-\frac{1-\nu}{\nu}} & ~~~( f_\sharp < f_{} < f^* )\cdots{\rm [TP]} \\
1 & ~~~( f^* < f_{} )\cdots{\rm [SF],\,[SS]} \\
\end{array} \right.~~~~
\label{density_F} \\
{\cal Z} (f_{})
&\simeq& \left\{ \begin{array}{ll}
f_{}^{z-2} & ~~~( f_\sharp < f_{} < f^* )\cdots{\rm [TP]} \\
f_{} & ~~~( f^* < f_{} )\cdots{\rm [SF],\,[SS]} \\
\end{array} \right.~~~~
\label{Z_f}
\end{eqnarray}
where $\sigma_F (f)$ is the line segment density at the fore end (pore)
and $f_\sharp \simeq N_0^{-\nu}$, $f^* \simeq 1$, and $f^{**} \simeq N_0^\nu$ are the characteristic forces separating different regimes.
The derivation of these forces and characteristic regimes with deformed shapes [TP], [SF], and [SS] are described in Ref.~\cite{EPJE_Saito_Sakaue_2011}.
The deformation characteristics illustrated in Fig.~2 are briefly introduced below:
(a) In the trumpet [TP] regime, the nonuniform deformation is analyzed in terms of the space-dependent blob model;
(b) In the stem-flower [SF] regime, the front end is almost fully stretched, whereas the rear end retains a blob-like shape;
(c) All the tensed segments are almost completely stretched in the strong-stretching [SS] regime.
Checking the Peclet number, we find that $Pe (R_{\rm eq}) \gsim 1 \Leftrightarrow f \gsim N_0^{-\nu}$, which covers the deformed shape regimes.

\begin{figure}[t]
\begin{center}
\includegraphics[scale=0.50]{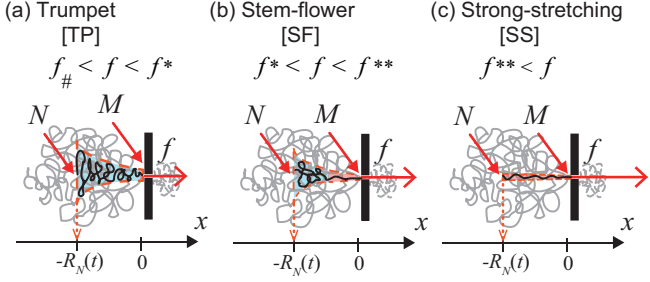}
      \caption{(Color online) Illustration of the deformed shapes associated with the different driving force magnitude. (a) Trumpet ($f_{\sharp} < f_{} < f^*$); (b) stem-flower ($f^* < f_{} < f^{**}$); (c) strong-stretching ($f^{**}<f_{}$). (See Ref.~\cite{EPJE_Saito_Sakaue_2011} for more details.) }
\label{fig2}
\end{center}
\end{figure}

\subsection{Stochastic Evolution of Tension Propagation}

In this section, we discuss the stochastic time evolution of tension propagation. 
By putting the segment flux at pore ${\rm d}M/{\rm d}t =\sigma_F (f)v_F$ into the time derivative of eq.~(\ref{N_M_f_R}), we have
\begin{eqnarray}
\frac{{\rm d}N}{{\rm d}t} = \sigma_F v_F + \sigma_F \frac{{\rm d}R_N}{{\rm d}t}
\label{time_evol_rigorous}
\end{eqnarray}
where $v_F=v(x=0)$ is the segment velocity at the pore.
To proceed, we apply the steady-state approximation by setting $V \equiv v_F$;
\begin{eqnarray}
\frac{{\rm d}N}{{\rm d}t} = \sigma_F V + \sigma_F \frac{{\rm d}R_N}{{\rm d}t}.
\label{time_evol_translocation}
\end{eqnarray} 
Note that the adopted steady-state ansatz $V \equiv v_F$ is different from that in our previous studies $V \equiv v_R =v(x=-R_N)$~\cite{PRE_Sakaue_2007,PRE_Sakaue_2010,EPJE_Saito_Sakaue_2011}. The reason for the present choice lies in the fact that it is consistent with the iso-flux condition required for the moving domain in the translocation process~\cite{JPCB_Rowghanian_Grosberg_2010}, while the previous one is not (see Appendix A).

Equation~(\ref{time_evol_translocation}) can be arranged into a stochastic differential equation that expresses the stochastic evolution of $t$ as a function of $N$ 
\begin{equation}
 {\rm d}t =  {\cal A} \, {\rm d}N + {\cal B} \, {\rm d}R_N
 \label{mass_translocation}
\end{equation}
\begin{eqnarray}
 {\cal A} &=& 1/(\sigma_F V) 
\label{translocation_A} \\
 {\cal B} &=& - 1/V 
 \label{translocation_B}
\end{eqnarray}
where ${\rm d}R_N$ is the noise originating from the initial random configuration.

At first sight, eq.~(\ref{mass_translocation}) seems to consist of a deterministic drift term and a noise term. However, this is not the case, since $V$ is also a stochastic variables.
By substituting eq.~(\ref{R_V_general})
into eq.~(\ref{mass_translocation}), we obtain
\begin{equation}
{\rm d}t  = (R_N/(\sigma_F{\cal Z})) \,{\rm d}N - (R_N/{\cal Z}) \,{\rm d}R_N 
\label{mass_translocation_03}
\end{equation}
Integrating this equation with respect to $N$ yields
\begin{equation}
 t  = (\sigma_F{\cal Z})^{-1} \int_0^{N} R_k {\rm d}k - (2{\cal Z})^{-1} R_N^2,
\end{equation}
where we adopt a midpoint discretization scheme in the definition of the stochastic integral\footnote{
We adopt the Stratonovich (midpoint discretization) scheme because it is expected to be suitable for describing real physical phenomena (random coil configuration) in terms of idealized delta-correlated noise and because it is compatible with conventional formulas in real analysis~\cite{SekimotoBook}.
Note, however, that the scaling exponents are not altered in this case, even if the Ito integral is employed.}. 
The second term is obtained from $\int_0^N R_k {\rm d}R_k \equiv \sum_{i=1}^N  (1/2)(R_k +R_{k-1}) [R_k -R_{k-1}] = (1/2)[R_k^2-R_0^2]$ and the initial condition $R_0(t=0)=0$.
This leads to the mean value:
\begin{eqnarray}
\langle t \rangle&=& (\sigma_F{\cal Z})^{-1} \int_0^N \langle R_k \rangle {\rm d}k - (2{\cal Z})^{-1} \langle R_N^2 \rangle 
\label{exact_mean_t_translocation} \\
				  &\simeq& \Lambda_1(\lambda=1)  (\sigma_F{\cal Z})^{-1} N^{1+\nu} -  {\cal Z}^{-1} N^{2\nu},~~  
\label{translocation_mean} 
\end{eqnarray}
where $\langle \cdots \rangle$ denotes the ensemble average over the initial conditions.
To derive eq.~(\ref{translocation_mean}), we assume that the spatial distance from the pore site to the $k$th segment's position at the initial time ($t=0$) has a normalized distribution function~\cite{deGennesBook,DoiBook}:
\begin{equation}
P_1 (R_k) = k^{-\nu} \psi_1 \left( R_k / k^{\nu} \right).
\label{dist_func_1body_translocation}
\end{equation}
This gives the mean value of the positive power of $R_k$ as
\begin{equation}
\langle R_k^\lambda \rangle = \int_{0}^{\infty} R_k^\lambda k^{-\nu} \psi_1 \left( \frac{R_k}{ k^{\nu} } \right) {\rm d}R_k = \Lambda_1 (\lambda) k^{\lambda \nu},
\label{translocation_1body_dist}
\end{equation}
where the numerical coefficient $\Lambda_1 (\lambda) \equiv \int_{0}^{\infty} u^\lambda \psi_1 \left( u \right) {\rm d}u $ follows from the variable transformation $u \equiv R_k/k^\nu$. 
The fact that $\psi_1 (u)$ is expected to decrease exponentially at the large $u$ guarantees the convergence of the above integral for positive $\lambda$.
We then substitute eq.~(\ref{translocation_1body_dist}) into eq.~(\ref{exact_mean_t_translocation}) because the exponents of all terms in eq.~(\ref{exact_mean_t_translocation}) are positive. 
For a strong force $f >1$ or a long chain $N \gg 1$, the first term in eq.~(\ref{translocation_mean}) is dominant, so we obtain the following scaling for the mean translocation time 
\begin{eqnarray}
\langle t \rangle 
&\simeq& \frac{N^{1+\nu}}{\sigma_F(f){\cal Z}(f)}  
\nonumber \\
&\simeq& 
\left\{ 
\begin{array}{ll}
N_0^{1+\nu} f^{1+\frac{1}{\nu}-z} & ~~~( f_\sharp < f_{} < f^* )\cdots{\rm [TP]} \\
N_0^{1+\nu} f^{-1} & ~~~( f^* < f_{} )\cdots{\rm [SF],\,[SS]} \\
\end{array} \right.~~~~.
\label{translocation_mean_dominant} 
\end{eqnarray}

The mean square time is given by
\begin{eqnarray}
\langle t^2 \rangle 
&=& (\sigma_F{\cal Z})^{-2} \langle \left[ \int_0^N R_k {\rm d} k \right]^2 \rangle + (2{\cal Z})^{-2} \langle R_N^4 \rangle \nonumber  \\
& &  - \sigma_F^{-1}{\cal Z}^{-2} \langle \left[ \int_0^N R_k {\rm d} k \right] R_N^{2} \rangle   
\label{exact_variance_t_translocation}   \\
&\simeq& \Lambda_2 (1,1) \, \sigma_F^{-2}{\cal Z}^{-2} N^{2+2\nu} +\Lambda_1(4) \, {\cal Z}^{-2} N^{4\nu} \nonumber \\
&& \qquad~ + \Lambda_2(1,2) \, \sigma_F^{-1}{\cal Z}^{-2} N^{1+3\nu},
\label{translocation_msq}
\end{eqnarray}
where, as before, we assume a normalized joint distribution function $P_2 (R_k,R_{k'}) = k^{-\nu} k'^{-\nu} \psi_2 \left( R_k /k^{\nu}, R_{k'}/ k'^{\nu}  \right)$ 
for the spatial distances $R_k$ and $R_{k'}$ at the initial time ($t=0$). 
This enables the mean value of the product of $R_k^{\lambda}$ and $R_{k'}^{\lambda'}$ to be calculated:
\begin{equation}
\langle R_k^\lambda R_{k'}^{\lambda'} \rangle = \Lambda_2 (\lambda,\lambda') k^{\lambda \nu} {k'}^{\lambda' \nu},
\label{dist_func_2body_translocation}
\end{equation}
where the numerical coefficient $\Lambda_2 (\lambda,\lambda') \equiv \int_{0}^{\infty} {\rm d}u \int_{0}^{\infty} {\rm d}u' u^\lambda u'^{\lambda'} \psi_2 \left( u,u' \right)$ follows from the variable transformation $u=R_k/k^\nu$ and $u'=R_{k'}/k'^\nu$.
Combining eqs.~(\ref{translocation_mean}) and (\ref{exact_variance_t_translocation}) gives the standard deviation ${\rm SD}(t) = \sqrt{ \langle t^2 \rangle - \langle t \rangle^2 }$. 
Since the cross-correlation $\langle \Delta R_k^\lambda  \Delta R_{k'}^\lambda \rangle = [ \Lambda_2 (\lambda,\lambda)  -\Lambda_1 (\lambda)^2 ]k^{\lambda \nu} {k'}^{\lambda \nu}>0$, where $\Delta R_k^\lambda \equiv R_k^\lambda -\langle R_k^\lambda \rangle$,
we obtain the following scaling relation for the standard deviation
\begin{eqnarray}
{\rm SD}(t) \simeq \frac{N^{1+\nu}}{\sigma_F(f){\cal Z}(f)},
\label{translocation_msq_2}
\end{eqnarray}
which has the same scaling structure as that of the mean time; i.e., ${\rm SD} (t) \sim \langle t \rangle$.

There are two distinct stages in the driven translocation process: tension propagation and post-propagation.
The above-mentioned scenario considers the tension-propagation stage only.
In the post-propagation stage, the moving domain consists of the entire chain on the {\it cis} side~\cite{EPJE_Saito_Sakaue_2011,PRE_Sakaue_2007,PRE_Sakaue_2010}. 
The total translocation time is then given by $\tau = \tau_{\rm p} + \tau_{\rm pp}$, where the first and second terms on the right-hand side respectively correspond to the tension-propagation and post-propagation periods. 
Equation~(\ref{time_evol_translocation}) can be applied to the post-propagation stage ($N=N_0 = const.$) dynamics by using the trivial condition ${\rm d} N_0/{\rm d} t =0$.
Combining it with eq.~(\ref{R_V_general}) leads to
\begin{equation}
{\rm d}t = -V^{-1} {\rm d} R_{N_0} \simeq -{\cal Z}(f)^{-1} R_{N_0}{\rm d} R_{N_0},
\label{diff_eq_pp}
\end{equation}
where it is noted that the stochasticity in the tension propagation pathway is no longer relevant, thus, eq.~(\ref{diff_eq_pp}) is a deterministic equation. 
Solving this, we find the post-propagation time $\langle \tau_{\rm pp} \rangle ={\cal Z} (f_{})^{-1} \langle {R_{N_0} (\tau_{\rm p}) }^2 \rangle \simeq {\cal Z} (f_{})^{-1} {N_0}^{2\nu}$.
 This means that, at the scaling level, $\tau_{\rm pp}$ is a correction term, so that the translocation time is eventually written as $\tau \simeq \tau_{\rm p}$~\cite{EPJE_Saito_Sakaue_2011,PRE_Sakaue_2007,PRE_Sakaue_2010}. In this study, we regard the translocation time as $\tau \simeq \tau_{\rm p}$.

\subsection{Discussion}

\begin{figure}[t]
\begin{center}
\includegraphics[scale=0.15]{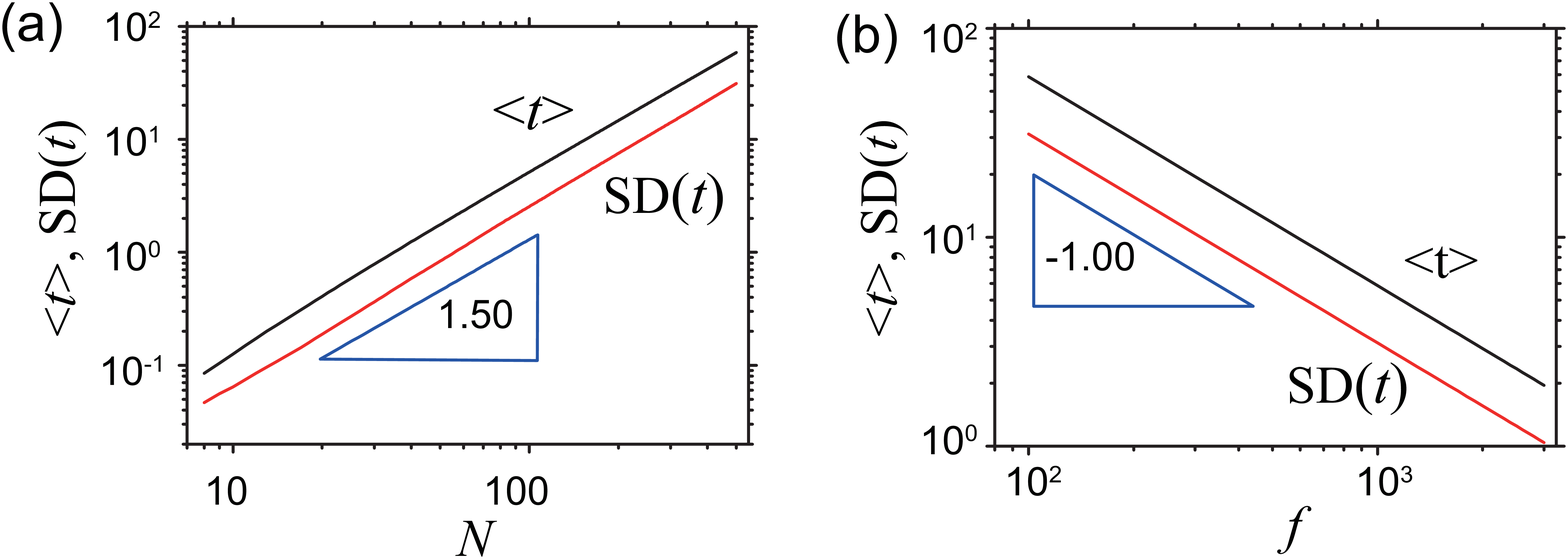}
      \caption{ Numerical calculation of the stochastic translocation process under the SF or SS regimes. Double logarithmic plots of the mean and standard deviation against (a) the segment number $N$ and (b) the magnitude of the driving force $f_{}$.}
\label{fig3}
\end{center}
\end{figure}
\begin{figure}[t]
\begin{center}
\includegraphics[scale=0.15]{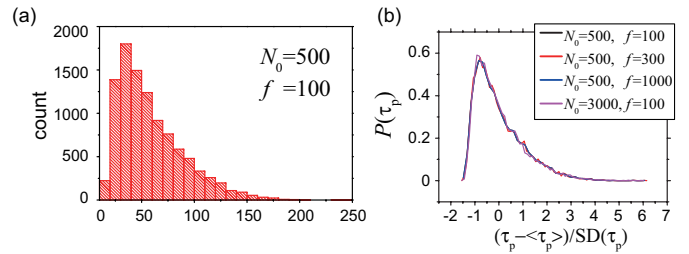}
      \caption{
	  (a) Frequency distribution of the process (tension-propagation) time $\tau_{\rm p} =t(N_0,f_{})~(\simeq \tau)$ for translocation with $N_0=500$, $f_{}=100$. (b) Probability density function as a function of normalized propagation time $(\tau_{\rm p} -\langle \tau_{\rm p} \rangle)/{\rm SD}(\tau_{\rm p})$ for various conditions. The SF or SS regimes are adopted and the sample number is $10,001$.}
\label{fig4}
\end{center}
\end{figure}

We have argued that the scaling exponents of the mean time are the same as those of the standard deviation in their respective regimes.
To test the above scaling predictions, we numerically integrated the stochastic differential equation (eq.~(\ref{mass_translocation_03})) for a Rouse chain (i.e., no excluded volume $\nu = 1/2$ and no hydrodynamic interactions $z=4$).
In this simplest case, the apparent random force ${\rm d}R_N/{\rm d}N$ becomes white noise~\cite{GrosbergBook} as 
$\langle \frac{{\rm d}R_N}{{\rm d}N} \frac{{\rm d}R_{N'}}{{\rm d}N'} \rangle = \delta (N-N')$, which enable us to treat the noise term in the stochastic differential equation as the Wiener process~\cite{GardinerBook}.
Such noise is generated by a reflecting boundary ($x=0$) that the propagation front cannot cross.
Figures 3--5 show the corresponding numerical results in the SF or SS regimes (eq.~(\ref{R_V_general}) and bottom lines of eqs.~(\ref{density_F}) and (\ref{Z_f})).
The triangles in Fig.~3 indicate the theoretical exponents;
they show that the slopes correspond very well with the numerical results.

The distribution function contains more information about the translocation time.
The profiles obtained in most numerical studies are asymmetric having a single left-skewed peak~\cite{EPJE_Bhattacharya_Luo_2009,JCP_Luo_2006,PRE_Bhattacharya_Binder_2010}. 
Rapid translocation experiments have also typically exhibited asymmetric profiles~\cite{PRE_Storm_2005,BJ_Wanunu_2008}, although some have given right-skewed or symmetric profiles~\cite{PRL_Meller_2001,NanoLett_Fologea_2005}.
Figure~4\,(a) demonstrates a histogram of the translocation time for the chain length $N_0=500$ obtained by our numerical integration. A characteristic profile with a single left-skewed peak in the time distribution is clearly observed. 
Similar asymmetry was observed over a broad range of parameters, as shown in Fig.~4\,(b) (where the translocation time is normalized as $(t -\langle t \rangle)/{\rm SD}(t)$). 
To gain a better understanding of the asymmetry, the following skewness ($=$third cumulant$/{\rm SD}^3$) was analyzed:
\begin{eqnarray}
\gamma_1 &\equiv& \langle \left[ \frac{t-\langle t \rangle}{{\rm SD}(t)} \right]^3 \rangle \nonumber \\
			&=& \frac{ \langle t^3 \rangle -3 \langle t^2 \rangle \langle t \rangle +2 \langle t \rangle^3 }{{\rm SD}(t)^3}.
\end{eqnarray}
Introducing the normalized joint distribution for the three segments $P_3(R_k,R_{k'},R_{k''}) = k^{-\nu} k'^{-\nu} k''^{-\nu} \psi_3 (u,u',u'')$ with $u=R_{k}/k^{\nu}$, $u'=R_{k'}/k'^{\nu}$, and $u''=R_{k''}/k''^{\nu}$, we obtain $\langle t^3 \rangle \sim \langle t \rangle^3$ through a similar discussion of the derivation of $\langle t \rangle$ and $\langle t^2 \rangle$. Thus, the skewness $\gamma_1$ is constant on the scaling level
and the profiles approximately overlap each other.
Recent numerical results~\cite{PRE_Bhattacharya_Binder_2010,EPJE_Bhattacharya_Luo_2009} are consistent with the present finding.

\begin{figure}[t]
\begin{center}
\includegraphics[scale=0.15]{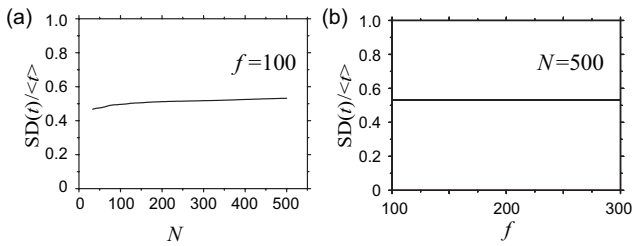}
      \caption{ Plots of the normalized standard deviation (NSD) ${\rm SD} (t)/\langle t \rangle$ against (a) the segment number $N$ and (b) the magnitude of the driving force $f_{}$ under the SF or SS regimes. }
\label{fig5}
\end{center}
\end{figure}

What factor is responsible for the asymmetric profile?
An equilibrium coil in free space retains a symmetric configuration. However, this is not the case for the initial configuration of a translocating polymer when a wall excludes the polymer, one end of which is located at the pore.
To clarify this point, we check the initial segment-distribution function obtained as follows:
\begin{equation} 
P_{\rm ini} (N,x)
=
\sqrt{2 N /\pi} e^{-x^2 /(2 N) },
\label{initial_conf}
\end{equation}
where $x \leq 0$ and $\int_{-\infty}^0 P_{\rm ini}(N,x) {\rm d}x = N$.
This is obtained by solving the diffusion equation 
$\partial_N P_{\rm ini} (N,x) = (1/2) \partial_x^2 P_{\rm ini} (N,x)$ with the reflecting boundary condition $\partial_x P_{\rm ini}(N,x)|_{x=0}=0$. 
In eq.~(\ref{initial_conf}), the reversal point $x_{\rm ref}$; i.e., $P_{\rm ini}(N,x_{\rm ref}+x) = P_{\rm ini}(N,x_{\rm ref}-x)$ for $x_{\rm ref} <0$ does not hold.
In other words, the configuration symmetry is violated; this is the only source of the asymmetric translocation time distribution in our model.
Since the tension-propagation mechanism under a strong force has received strong support from numerical studies~\cite{EPL_Lehtola_2009,PRE_Bhattacharya_Binder_2010,arXiv_Ikonen_Sung_2011}, and, according to Fyta~{\it et al.}~\cite{PRE_Fyta_2008}, the translocation time depends on the initial configuration, it is reasonable to assume that stochastic evolution of the tension propagation along the initial asymmetric configuration gives rise to the left-skewed profiles observed in numerical and experimental studies.
Conversely, the detection of other translocation time profiles (e.g., right-skewed and symmetric profiles~\cite{PRL_Meller_2001}) may be regarded as a good indicator that other factors (such as specific interactions in the pore) dominate the process.

Figures~5\,(a) and (b) show plots of the normalized standard deviation (NSD) ${\rm SD} (t) /\langle t \rangle$.
It is nearly constant against the parameters $f_{}$ and $N_0(\gg 1)$.
The experimental results in ref.~\cite{PRE_Storm_2005} collecting simple translocation events without folding show that the NSD
has values in the range $\simeq 0.1-0.2$ independent of the chain length; this is in qualitative agreement with our prediction.

\section{Stretching}

\begin{figure}[b]
\begin{center}
\includegraphics[scale=0.50]{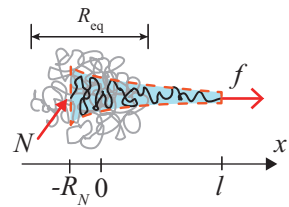}
      \caption{(Color online) Schematic representation of the stretching process (in the trumpet [TP] regime). The tensed moving domain is shaded. }
\label{fig6}
\end{center}
\end{figure}

In this section, we consider a different protocol for chain transport in which one end of a linear polymer chain is pulled suddenly, as shown in Fig.~6.
If the linear polymer chain is initially relaxed in solution and one end is pulled by a constant force $f_{}$ in the $x$-direction after $t=0$, 
the chain will be progressively deformed and will eventually settle in a steady state~\cite{Sakaue_Wada}.
Applying the two-phase picture and the stochastic method in a similar manner as above, we analyze the stretching dynamics and derive the mean time of the transient process and its standard deviation.
The segments are numbered from the pulled segment (one chain end) to the other end.
The positions of the front and the rear ends in the moving domain are $x=l(t)$ and $-R_N(t)$ respectively; the pulled segment is initially located at the origin.
Here again, we utilize the dynamical equation of state to characterize the global conformation~\cite{Sakaue_Wada}:
\begin{eqnarray}
N &=& \sigma_F (f_{}) L   
\label{N_L_dragged_eq_pull} \\
L V &=& {\cal Z}(f_{})  
\label{L_V_dragged_eq_pull} .
\end{eqnarray}
where the extension along the $x$-direction is $L= l+R_N$ (see Fig.~6), and $\sigma_F (f_{})$, ${\cal Z} (f_{})$ are given in eqs.~(\ref{density_F}),~(\ref{Z_f}), respectively. Evidently, eqs.~(\ref{N_L_dragged_eq_pull}) and~(\ref{L_V_dragged_eq_pull}) exactly corresonds to eqs.~(\ref{N_M_f_R}),~(\ref{R_V_general}), respectively.

\subsection{Stochastic Evolution of Tension Propagation}
Combining the boundary condition ${\rm d} l(t)/{\rm d}t = v_F(t) \simeq V(t)$ with the time derivative of $l = L - R_N$, we obtain
\begin{equation}
\frac{{\rm d}L}{{\rm d}t} = V + \frac{{\rm d}R_N}{{\rm d}t}.
\label{pull_dL_V_dR_N}
\end{equation}
Combining eqs.~(\ref{N_L_dragged_eq_pull}) and (\ref{L_V_dragged_eq_pull}) with eq.~(\ref{pull_dL_V_dR_N}) gives the stochastic equation:
\begin{equation}
{\rm d} t	= {\cal A}\,{\rm d} N  +{\cal B}\,{\rm d}  R_N
\label{dt_eq_pull}
\end{equation}
where ${\rm d}R_N$ acts as an apparent random force with $\langle {\rm d}R_N \rangle =0$, and
\begin{eqnarray}
{\cal A} &=& \sigma_F (f)^{-2} {\cal Z} (f)^{-1} N
\label{pull_A} \\
{\cal B} &=& - \sigma_F (f)^{-1} {\cal Z} (f)^{-1} N.
\label{pull_B}
\end{eqnarray}
In contrast to eqs.~(\ref{translocation_A}) and (\ref{translocation_B}) for the translocation dynamics, eqs.~(\ref{pull_A}) and (\ref{pull_B}) contain only deterministic variables.
Integrating eq.~(\ref{dt_eq_pull}) with respect to the segment number leads to
\begin{equation}
 t = \frac{ N^2 }{ 2 \sigma_F^2 {\cal Z} } - \frac{ 1 }{ \sigma_F {\cal Z} } \int_0^N k \frac{{\rm d} R_k}{{\rm d} k} {\rm d}k.
\label{t_pull}
\end{equation}
We then take the ensemble average to give
\begin{eqnarray}
\langle t \rangle &=&  \frac{N^2}{2 \sigma_F(f)^2 {\cal Z}(f) }
\nonumber \\
&\simeq & \left\{ 
\begin{array}{ll}
N^{2} f^{\frac{2}{\nu}-z} & ~~~( f_\sharp < f_{} < f^* )\cdots{\rm [TP]} \\
N^{2} f^{-1} & ~~~( f^* < f_{} )\cdots{\rm [SF],\,[SS]} \\
\end{array} \right.~~~~.
\label{ave_sudden_pull}
\end{eqnarray}
The mean square of the process time is calculated to be
\begin{eqnarray}
\langle t^2 \rangle 
&=& \frac{ N^4 }{ (2 \sigma_F^2 {\cal Z})^2 } + \frac{ 1 }{ \sigma_F^2 {\cal Z}^2 } \int_0^N {\rm d}k \int_0^N {\rm d}k' \, k k' \langle \frac{{\rm d}R_k}{{\rm d}k} \frac{{\rm d}R_{k'}}{{\rm d}k'} \rangle  \nonumber \\
&=& (2 \sigma_F^2 {\cal Z})^{-2} N^4 + c(\sigma_F {\cal Z} )^{-2}  N^{2+2\nu}, 
\end{eqnarray}
where $c$ is a numerical coefficient of order unity. Note that $\langle \frac{dR_k}{dk} \frac{dR_{k'}}{dk'} \rangle = \delta (k-k')$ and $\langle \frac{dR_k}{dk} \frac{dR_{k'}}{dk'} \rangle \simeq (k-k')^{2 \nu-2}$ for ideal and self-avoiding chains, respectively.
We then obtain the standard deviation
\begin{eqnarray}
{\rm SD} (t)&\simeq& \frac{ N^{1+\nu}}{ \sigma_F(f) {\cal Z}(f) },
\nonumber \\
&\simeq & \left\{ 
\begin{array}{ll}
N^{1+\nu} f^{1+\frac{1}{\nu}-z} & ~~~( f_\sharp < f_{} < f^* )\cdots{\rm [TP]} \\
N^{1+\nu} f^{-1} & ~~~( f^* < f_{} )\cdots{\rm [SF],\,[SS]} \\
\end{array} \right.~~~~.
\label{SD_sudden_pull}
\end{eqnarray}
and the NSD
\begin{eqnarray}
\frac{{\rm SD}(t)}{\langle t \rangle}  &\simeq& \sigma_F(f_{}) N^{-1+\nu}.
\label{NSD_sudden_pull}
\end{eqnarray}

\subsection{``Stretching" vs. ``Translocation"}
Comparison of eqs.~(\ref{translocation_mean_dominant}) and~(\ref{translocation_msq_2}) with eqs.~(\ref{ave_sudden_pull}) and~(\ref{SD_sudden_pull}) reveals that the process distributions for ``translocation and ``stretching" differ remarkably. 
In the former case, the scaling exponents of the mean and the standard deviation are identical so that ${\rm SD}(t)/\langle t \rangle \rightarrow {\rm const.}>0$ in the limit $N \rightarrow \infty$.
In the latter case, eq.~(\ref{NSD_sudden_pull}) indicates ${\rm SD}(t)/\langle t \rangle \rightarrow 0$ in the limit $N \rightarrow \infty$.
To determine the cause for this difference between the two cases, we compare the relevant equations by listing them again:

\begin{eqnarray} 
{\rm Translocation} \qquad \qquad
&\Leftrightarrow&
~~~{\rm Stretching} \nonumber  \\*
\begin{array}{ll}
~~~~~ N-M = \sigma_F(f)L \\
\qquad ~~~~ LV = {\cal Z}(f) \\
(N-M) V = \sigma_F(f){\cal Z}(f) \\
\end{array} 
\Biggr\}
&\Leftrightarrow& 
\Biggl\{ 
\begin{array}{ll}
~~N = \sigma_F(f)L \\
~LV = {\cal Z}(f) \\
N V = \sigma_F(f){\cal Z}(f) \\
\end{array} 
\label{dyn_eq_cmp}  \\*
\nonumber \\*
L=R_N \qquad \qquad ~~~
&\Leftrightarrow& L=R_N+l 
\label{extension_cmp}
\end{eqnarray}
Equations~(\ref{dyn_eq_cmp}) are a set of different forms in the dynamical equations of state, eqs.~(\ref{extension_cmp}) give the chain extension in the moving domain. 
The scaling forms of $\sigma_F(f)$ and ${\cal Z}(f)$ are given in eqs.~(\ref{density_F}) and~(\ref{Z_f}).
As is clear from this comparison, the essential difference is found in eq.~(\ref{extension_cmp}), i.e., in the translocation process, the segments with their label smaller than $M(t)$, which are in the trans-side, are already free from the driving tension, thus, are not relevant to the moving domain.

First, this leads to the so-called iso-flux condition $N(t) \simeq  M(t)$ to leading order~\cite{JPCB_Rowghanian_Grosberg_2010}. One can indeed check $N(t)  \gg \sigma_F \langle R_N(t) \rangle$ using the first moment relation $\langle R_N(t) \rangle \simeq N(t)^{\nu}$ (eq.~(\ref{translocation_1body_dist})). 

Second, this causes the qualitative difference in the stochastic effect in two processes.
In the stretching dynamics, the evolution of the moving domain has two components (Fig.~6, the right equation in eq.~(\ref{extension_cmp})).
One component is the pulled out part in the region $0 \leq x \leq l$ and the other component is the rear part ($-R_N\leq x \leq 0$) due to the tension spreading.
The former dominates the process ($ l \gg R_N$) giving rise to the deterministic term in the stochastic differential equations (eqs.~(\ref{dt_eq_pull}) and (\ref{pull_A})), while the latter acts as a small noise term (eqs.~(\ref{dt_eq_pull}) and (\ref{pull_B})).
In contrast, in the translocation dynamics, the moving domain has only a stochastic component $R$ (Fig.~1, the left equation in eq.~(\ref{extension_cmp})), which results in the non-vanishing NSD, even in the long-chain limit.

\section{Remarks and Summary}

Before concluding, let us make some comments concerning the relation between the present study and other theoretical studies of polymer translocation.

First, we have focused on the ``driven regime" (i.e., $f>f_{\sharp}\simeq N_0^{-\nu}$) for which the chain deformation dynamics are important alongside tension propagation.
In particular, as noted in Sec. II, we expect that our approach can accurately calculate the process time distribution in the strong force regime $Pe (a)>1 \Leftrightarrow f > f^*\simeq 1$.
In such situations, the retardation effect due to segment accumulation on the {\it trans} side provides only a weak perturbation so that it can be neglected for asymptotic scaling. 
However, it is essential to appropriately treat the {\it trans} side to describe the translocation process in the weak-force ``near-equilibrium" regime $f<f_\sharp$, as has been done in several studies~\cite{EPL_Dubbeldam_2007,JPhys_Vocks_Panja_2008,PRL_Sung_1996}.
In this regime, the Brownian force also greatly influences the process time distribution. 
It is interesting to determine the scaling structure of the standard deviation in such a weak-force regime.

Second, the pore has been assumed to be only geometric constrictions. 
We have restricted ourselves to this simplest situation to get a clear-cut impact of the initial polymer configuration on the process time distribution.
However, elucidating the case with functionalized pores with friction, confinement or specific interactions with polymers is an important problem.

To summarize, we have discussed the process time distribution for driven polymer transport inherent in a flexible molecular chain.
The tension propagates with time along random paths following the initial configuration and the stochastic propagation mechanism is introduced as the distribution origin.
We give the formulation for the two cases of translocation and stretching, which involve different forcing points.
The forcing point of the former process is fixed in the rest frame, whereas that of the latter process is fixed to the chain end.
Our analysis predicts the scaling exponents of the first and second cumulants (mean, variance) in the process time. 
In translocation dynamics, the probability distribution of the process time has a characteristic asymmetric shape, which may reflect the initial shapes, and fluctuation effects remain substantial even in the long-chain limit. 
On the other hand, the broadness of the distribution in stretching dynamics becomes unnoticeable in the long-chain limit.

~

\section*{Acknowledgements}

This work was supported in part by the JSPS Core-to-Core Program
``International research network for non-equilibrium dynamics of soft matter".
The authors are also grateful to an anonymous referee for his/her helpful comments on the relevance of the iso-flux condition to the translocation process.

\section*{APPENDIX A: STEADY STATE ANSATZ AND ISO-FLUX CONDITION}

Recently, Rowghanian and Grosberg pointed out a unique property inherent to the moving domain in the translocation process, that is to say, the segment flux is almost constant everywhere in the moving domain~\cite{JPCB_Rowghanian_Grosberg_2010}. This leads to the balance of the fluxes into and out of the moving domain, i.e., $\sigma_F v_F \simeq \sigma_R v_R$, where $v_R=v(x=-R_N)$ and $\sigma_R=\sigma(x=-R_N)$. 
The subscript ``$R$" indicates the rear part of the moving domain, where the last tensed blob of size $\xi_R \simeq v_R^{1/(1-z)}$ is located.
In the appendix, we shall make a connection between the iso-flux condition and our steady-state ansatz.

The steady state approximation would be valid under the condition;
$\tau_{\rm relax} \dot{\gamma} \lsim 1 \Leftrightarrow \delta v \lsim V$ with
the representative (or average) velocity $V(t)=v(x,t)$ with $\exists x \in [0,-R_N]$, the relaxation time $\tau_{\rm relax}\simeq R_N/V$ and the shear rate $\dot{\gamma} \simeq \delta v/R=(v_F- v_R)/R_N$.
For the concreteness, we focus on the TP regime and investigate the velocity difference inside the moving domain under iso-flux condition. The flux balance $\sigma_F v_F \simeq \sigma_R v_R$ is transformed as
\begin{eqnarray}
v_F/v_R=\sigma_R/\sigma_F \simeq (1-\delta v_R/V)^{-q} (fR)^q
\end{eqnarray}
where $v_F=V+\delta v_F$, $v_R=V-\delta v_R$ and $\sigma_R \simeq g_R/\xi_R \simeq v_R^{-q}$ with $q=(1-\nu)/[(z-1)\nu]$.
Then we find
\begin{eqnarray}
1+\frac{\delta v_0}{V} = (fR)^q \left( 1-\frac{\delta v_R}{V} \right)^{1-q}.
\label{iso-flux-}
\end{eqnarray}
Note that $(fR)^q>1$ holds under the deformed shape regimes. 

(i) If we assume $V(t)=v_R$, the iso-flux condition eq.~(\ref{iso-flux-}) becomes $\delta v/V =(fR)^q-1$, which might not satisfy the condition for the steady-state approximation.

(ii) If $V(t)=v_F$,  the iso-flux condition eq.~(\ref{iso-flux-}) becomes $\delta v/V=1-(fR)^{-\frac{q}{1-q}} <1$, thus the condition for the steady state approximation is satisfied. In a similar way, the validity of the steady-state approximation with $V=v_F$ is verified in the SF regime, too, while the iso-flux condition is trivially realized under the SS regime.

In addition, we can check the correspondence at the level of the dynamical equation.
The local force balance under TP regime is generally given by
\begin{equation}
\frac{1}{\xi(x,t)} = \int_{-R_N}^0 v(x,t) \xi(x,t)^{z-2} dx
\end{equation}
Its spatial derivative leads to $-\partial_x \xi(x,t)= v(x,t) \xi(x,t)^{z-1}$.
We find its solution as
\begin{equation}
\xi(x,t) = \left[ \int_{-R_N}^{x} v(x,t) dx \right]^{\frac{1}{2-z}}
\end{equation}
Putting the force balance at pore $f=1/\xi_F=1/\xi(x=0,t)$ leads to
\begin{eqnarray}
R_N \left[ \frac{1}{R_N} \int_{-R_N}^{0} v(x,t) dx \right] &=& {\cal Z}(f).
\label{d_eq_state_appendix}
\end{eqnarray}
This is the dynamical equation of state (eqs.~(\ref{R_V_general})) with the ``average velocity" $V(t) \equiv \frac{1}{R_N} \int_{-R_N}^{0} v(x,t) dx$.

In evaluating the average velocity $V(t)$, the fore end has the dominant weight under the iso-flux $J(t)$ condition, since $v(x,t)=J(t)/\sigma(x,t)$ increases towards the fore part given the nonuniform spatial segment profile.
Thus, $V(t)=\left[ \frac{1}{R_N} \int_{-R_N}^{0} v(x,t) dx \right] \simeq v(x=0,t) = J(t)/\sigma(x=0,t)$ is expected so that the reasonable correspondence between the steady state approximation with the ansatz $V(t)=v_F(t)$ and the iso-flux model~\cite{JPCB_Rowghanian_Grosberg_2010} is found.
By multiplying $\sigma_F$ to both sides of eq.~(\ref{R_V_general}), we obtain the dynamical equation of state in terms of the flux $J(t)$ as $R_N J = \sigma_F(f) {\cal Z}(f)$, which coincides with the result based on the iso-flux model~\cite{JPCB_Rowghanian_Grosberg_2010}.

\end{document}